\documentclass[onecolumn,letterpaper,11pt,draftclsnofoot]{IEEEtran}
\usepackage{amssymb}
\usepackage{amsmath}
\usepackage{amsthm}
\usepackage{amsfonts}
\usepackage{mathrsfs}
\usepackage{float}
\usepackage{graphicx}
\usepackage{epsfig,graphics,subfigure,psfrag}
\usepackage{pdfpages}
\usepackage{enumerate}
\usepackage{engord}
\usepackage{cite}
\usepackage{verbatim}
\usepackage[ruled]{algorithm2e}

\begin{document}

\newtheorem{definition}{\bf Definition}
\newtheorem{theorem}{\bf Theorem}
\newtheorem{lemma}{\bf Lamma}
\newtheorem{proposition}{\bf Proposition}
\newcommand{\titlefontsize}{\fontsize{19pt}{20}\selectfont}

\title{\titlefontsize UAV Aided Aerial-Ground IoT for Air Quality Sensing in \\ Smart City: Architecture, Technologies and Implementation}

\author{
\IEEEauthorblockN{
\normalsize{Zhiwen Hu},
\normalsize{Zixuan Bai},
\normalsize{Yuzhe Yang},
\normalsize{Zijie Zheng},
\normalsize{Kaigui Bian},
and
\normalsize{Lingyang Song}
 \\}
\IEEEauthorblockA{\normalsize{School of Electronics Engineering and Computer Science\\ Peking University, Beijing, China\\
Email: \{zhiwen.hu, zixuan.bai, yuzhe.yang, zijie.zheng, bkg, lingyang.song\}@pku.edu.cn} \\
}
}
\maketitle

\thispagestyle{empty}
\begin{abstract}
As air pollution is becoming the largest environmental health risk, the monitoring of air quality has drawn much attention in both theoretical studies and practical implementations.
In this article, we present a real-time, fine-grained and power-efficient air quality monitoring system based on aerial and ground sensing.
The architecture of this system consists of four layers: the sensing layer to collect data, the transmission layer to enable bidirectional communications, the processing layer to analyze and process the data, and the presentation layer to provide graphic interface for users.
Three major techniques are investigated in our implementation, given by the data processing, the deployment strategy and the power control.
For data processing, spacial fitting and short-term prediction are performed to eliminate the influences of the incomplete measurement and the latency of data uploading.
The deployment strategies of ground sensing and aerial sensing are investigated to improve the quality of the collected data.
The power control is further considered to balance between power consumption and data accuracy.
Our implementation has been deployed in Peking University and Xidian University since February 2018, and has collected about 100 thousand effective data samples by June 2018.
\end{abstract}

\section{Introduction}

Based on a recent report from the World Health Organization~\cite{bib_WHO}, air pollution has become the greatest threat to human health, which is responsible for one in eight of deaths each year.
The contaminants in the air are mostly caused by the exhaust emission from industrial production procedures and the daily activities of residents, such as driving fuel automobiles or incinerating garbages~\cite{bib_AriPollution}.
To quantitatively describe the degree of air pollution, the governments define air quality index (AQI), which depends on the concentrations of some air pollutants, including the fine Particulate Matters (e.g., PM$_{2.5}$ and PM$_{10}$) and other chemical substances~\cite{bib_AQI}.
A higher concentration of a particular pollutant leads to a higher value of AQI, indicating a more severe air pollution.
To measure the concentration of a specific pollutant, large professional instruments with high precision or tiny commercial sensors with low cost can be utilized~\cite{bib_Sensing}.

Currently, the most authoritative monitoring systems belong to the official meteorological monitoring bureaus of the governments.
Despite the high precision they can achieve, these systems only set up a few observation stations over a large area and provide measurement results with significant delay~\cite{bib_Station}.
Recent studies show that air quality has the intrinsic characteristic to change from meters to meters, especially in the urban areas with complicated terrain that caused by densely distributed tall buildings~\cite{bib_Meter2Meter}.
Therefore, it is preferred that the fine-grained and real-time air quality sensing is performed based on a large number of tiny sensors, which can be densely deployed and have faster sensing procedure.
Such solution creates a special application of Internet-of-Things (IoT) in smart city~\cite{bib_IoTAir}, where massive data can be collected and analyzed~\cite{bib_IoTData}.
In this way, an air quality distribution map with high spatial resolution and low latency can be established to provide valuable information for citizens, such as suggesting them to keep away from highly polluted area or helping decide the best ventilation system for a building~\cite{bib_Suggestions}.

In this article, we present our aerial-ground air quality sensing system in smart city, which has a four-layer architecture, including the sensing layer, the transmission layer, the processing layer, and the presentation layer.
The sensing layer mainly consists of IoT-based sensing devices that perform ground sensing and aerial sensing.
The transmission layer is based on the existing wireless communication network, which enables bidirectional communications.
The processing layer is responsible for controlling the sensing devices, as well as receiving, recording and analysing data for further usage.
The presentation layer has the graphic user interface~(GUI) to provide the real-time map of 3D air quality distribution.

To make it a state-of-the-art application of IoT technology in smart city, this sensing system is also designed to be fine-grained, real-time and power-efficient at the same time.
To achieve these targets, we investigate three important techniques in our implementation.

The first and the most significant technique resides in \emph{data processing}.
Specifically, the server in the processing layer is able to perform spatial fitting and short-term prediction.
Spacial fitting indicates the usage of the collected data from measured locations to provide a best guess of the air quality at the unmeasured locations.
Short-term prediction represents the usage of the historical data and other weather conditions to deduce the 3D air quality distribution in the near future.
In this way, the influence of the incomplete measured values and the affect of the data uploading latency can be eliminated.

The second technique is the \emph{deployment strategy}.
As there are limited number of sensing devices on the ground and limited power of the UAV to detect wide-range 3D space, the deployment of ground sensing and aerial sensing could influence the effectiveness of the spatial fitting and short-term prediction.
For ground sensing, the locations of the fixed sensing devices should be properly selected, so as to establish a reasonable distribution of monitoring points.
For aerial sensing, the aerial sensing positions and routing scheme should be carefully considered, so as to improve the completeness of the spatial detection.

The third and last technique is \emph{power control}.
The sensors in smart city usually has no external power supply, only with a battery with limited electricity.
In our sensing system, there is an intrinsic tradeoff between the power consumption and the precision of spatial fitting and short-term prediction.
A more frequent data collection procedure can provide a better continuity and a better precision, but it also induces additional power consumption.
In our implementation, a proper power control scheme is designed to balance between the power consumption and the precision.

The rest of this article is organized as follows:
The four-layer architecture is presented in details in Section~\ref{sec_Architecture}.
The data processing as the key technique is provided in Section~\ref{sec_Processing}.
The deployment strategies of ground and aerial sensing are given in Section~\ref{sec_Deployment}.
The power control to balance between precision and power consumption is discussed in Section~\ref{sec_Power}.
Finally, the conclusion is drawn in Section~\ref{sec_Conclusion}.

\section{System Architecture}\label{sec_Architecture}

\begin{figure}[!thp]
\centering
\includegraphics[width=5in]{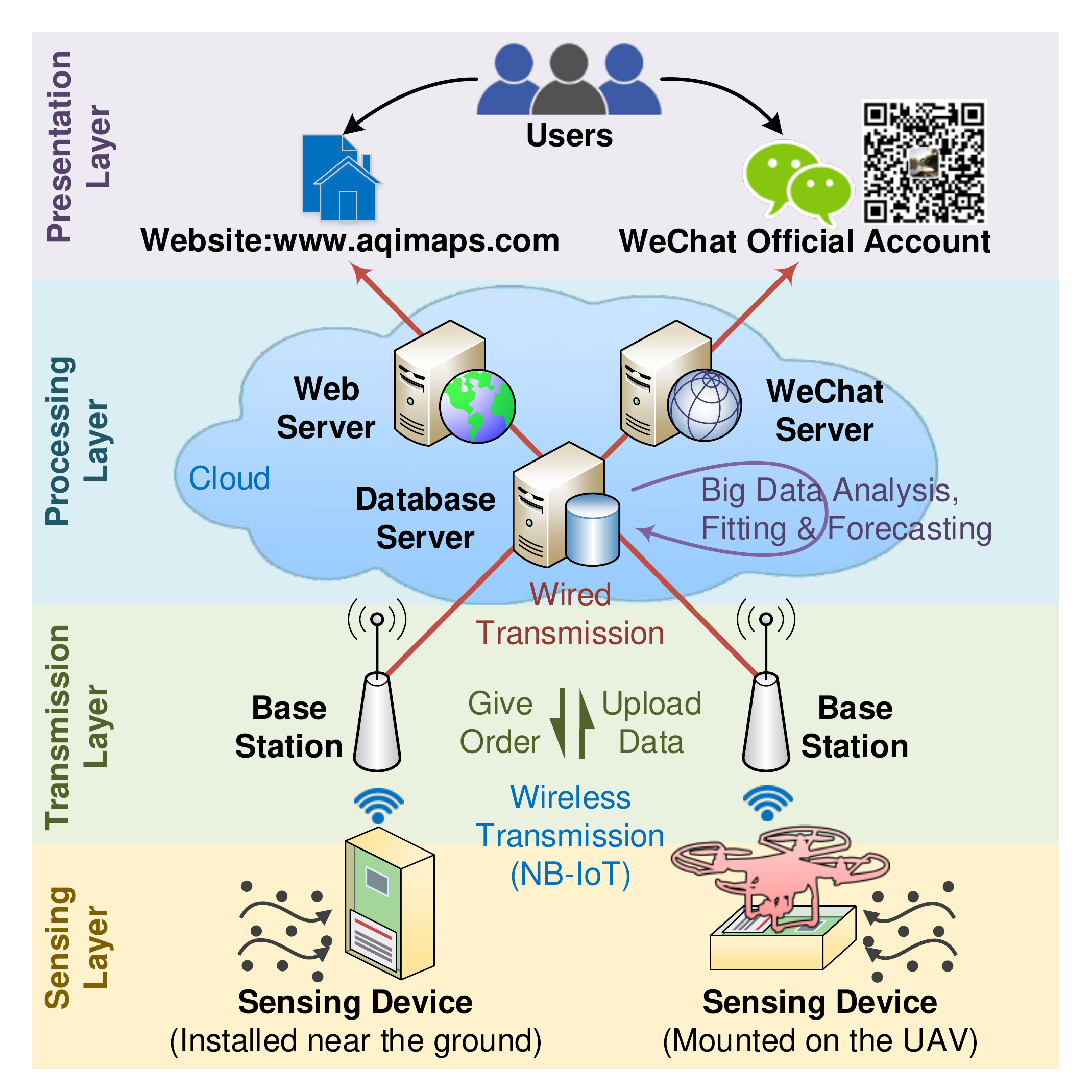}
\vspace{-5mm}
\caption{The architecture of our system, which mainly consists of four layers: the sensing layer, the transmission layer, the processing layer and the presentation layer.}\label{Fig_1_Architecture}
\end{figure}

As shown in Fig.~\ref{Fig_1_Architecture}, our system consists of four layers, namely, the transmission layer, the processing layer and the presentation layer.
The function of the sensing layer is to collect the data of real-time air quality.
It is carried out by the sensing devices, either installed near the ground or mounted on a mobile UAV.
The function of the transmission layer is to enable the bidirectional communications between the sensing layer and the processing layer.
It is supported by the infrastructure of the current wireless communication networks.
The function of the processing layer is to record the data from the sensing layer and submit the processing results to the presentation layer.
It is implemented in the cloud server where big data analysis and real-time processing are supported.
The function of the presentation layer is to provide valuable information for the users.
It includes our official website and our official WeChat subscription account.

In the following four subsections, we introduce the design considerations and the implementation overview of these layers.

\subsection{Sensing Layer}

The basic components of the sensing layer are the sensing devices.
These devices are responsible for collecting data from the environment and transmitting the data to the processing layer.
To guarantee a long battery duration without external power supply, these devices should be carefully designed and programmed.
Their composition, function, and deployment are presented as follows:

\subsubsection{Composition}\emph{}\!\!
In general, a proper air quality sensing device should at least contain: an air quality sensor to generate data based on the environment, a Micro-Controller Unit (MCU) to control the task logic of the device, a wireless transmission module to enable data transmission, and a battery with enough capacity to support long-term missions.
In our implementation, without the loss of generality, we mainly focus on the value of PM$_{2.5}$. The adopted sensor is A3-IG, with a small size and a quick response to the change of the environment.
ATmega128A is chosen as the MCU, and SIM7000C is chosen as the wireless communication module, all of which have a low power consumption.
Each device has a 13600mAh rechargeable battery, as a proper choice to balance between duration and weight.

\subsubsection{Function}\emph{}\!\!
The ultimate goal of each sensing device is to detect the quality of air, and generate the corresponding data in the sequence of time.
Two times of detections are not necessarily to be very close, since each detection consumes a certain amount of power.
In addition, uploading data to the processing layer also causes power consumption.
Therefore, in our implementation, detecting intervals and uploading intervals of different devices are controlled by the server in the processing layer.

\subsubsection{Deployment}\emph{}\!\!
Both ground sensing and aerial sensing are necessary to provide the 3D fine-grained profiling.
In our implementation, the devices that execute ground sensing are fixed near the ground, and the devices that execute aerial sensing are tied beneath the UAVs.
Ground sensing is able to continuously collect data in the long run, but the air quality off the ground cannot be detected, which is quite important for a dense-urban area.
Aerial sensing, on the contrary, is able to sense the air quality off the ground, but it cannot be executed continuously due to the high consumption of battery power and human labor.
By combining these two kinds of sensing schemes, a better set of data can be collected.
The server is able to utilize spatial fitting and short-term prediction to calculate the data at the time when no aerial sensing is performed, which is explained in Section~\ref{sec_Processing}.
Our system has been deployed in Peking University and Xidian University since Feb.~2018, with 60 outdoor devices and 120 indoor devices, as shown in Fig.~\ref{Fig_2_HardwareAndSoftware}~(a).
It has collected almost 100 thousand effective data samples by June 2018. Each sample contains the sensing location, the sensing time, the detected PM$_{2.5}$ value and the detected PM$_{10}$ value~\cite{bib_DataSet}.

\subsection{Transmission Layer}

The transmission layer is responsible for the bidirectional communications between the server in the processing layer and the sensing devices in the sensing layer.
It can be considered as the physical infrastructure that enables the transmission and the logical transmission protocol that specifies the method of communications.

\subsubsection{Physical Infrastructure}\emph{}\!\!
Based on different solutions, the physical infrastructure in the transmission layer could be different.
If the processing layer is connected to the Internet and the sensing layer transmits data in a wireless way, then the base stations or the WiFi access points play the roles of the physical infrastructures.
In some of the solutions where ad-hoc is used to aggregate sensing data~\cite{bib_AdHoc}, the sensing devices could also play the roles of the transmission infrastructures.
In our implementation, the amount of each time's transmitted data is small, and a certain amount of latency is tolerable.
Thus the 4G base stations that support Narrow Band IoT (NB-IoT) transmission are considered as the physical infrastructure.

\subsubsection{Logical Protocol}\emph{}\!\!
The design of the logical protocol influences the behaviour of the sensing devices and the server.
In the air quality sensing system, such protocol specifies when and how the data is uploaded to the server or the command is downloaded by the sensing devices.
The design of the logical protocol greatly influences the power consumption, since the sensing devices are expected to sleep during idle periods and cannot receive the command from the server until it wakes up.
In our implementation, when the sensor is awake and connected to the base station, it uploads data to the server by sending these data to the static IP address of the server.
To adjust the setting of a specific device, the server can put the command in the response information to the device just after the reception of the uploaded data, or directly send a short message which can only be noticed by the device as it wakes up.
Each device is waken up by the external timer interrupt once a minute, to check whether it is time to detect air, to upload data or to sleep again.

\subsection{Processing Layer}

This layer is responsible for receiving and recording the data transmitted from the sensing layer.
In addition, data processing and decision making are also implemented in this layer.
In the following, we briefly discuss the potential usage of data processing and decision making, respectively.

\subsubsection{Data Processing}\emph{}\!\!
First, the server should pre-process the raw data received from the sensing devices, in which way the sensing results are calibrated and the abnormal values are eliminated.
Second, the server should utilize the limited number of collected data to infer the air quality at undetected locations.
Such technique can be considered as spacial fitting, where a model of the air quality distribution should be trained and established in advanced.
Third, the server should be able to provide a prediction of the air quality in the near future if possible.
The ability of prediction can provide a more timely result to build the air quality map in real-time against the latency of data uploading.
In addition, the users can also refer to the prediction of air quality so as to adjust their daily plans in advance.

\subsubsection{Decision Making}\emph{}\!\!
Based on the processed data and the trained models, the processing layer is able to generate some decisions.
These decisions are able to optimize the system from multiple considerations.
The detecting interval and the uploading interval can be adjusted at any time to balance between the precision and the power efficiency.
These settings are directly transmitted to the sensing devices and take effect immediately.
The locations of the ground sensing devices and the flight route of the UAV can also be adjusted to collect better data for a more accurate distribution model.
The executions of these plans have to be done manually.

\subsection{Presentation Layer}

This layer provides users with GUI to realize information interaction.
In addition to the interface that facilitates the query from ordinary users, the interface reserved for the system manager is also important.

\subsubsection{GUI for Ordinary Users}
\begin{figure}[!thp]
\centering
\includegraphics[width=5in]{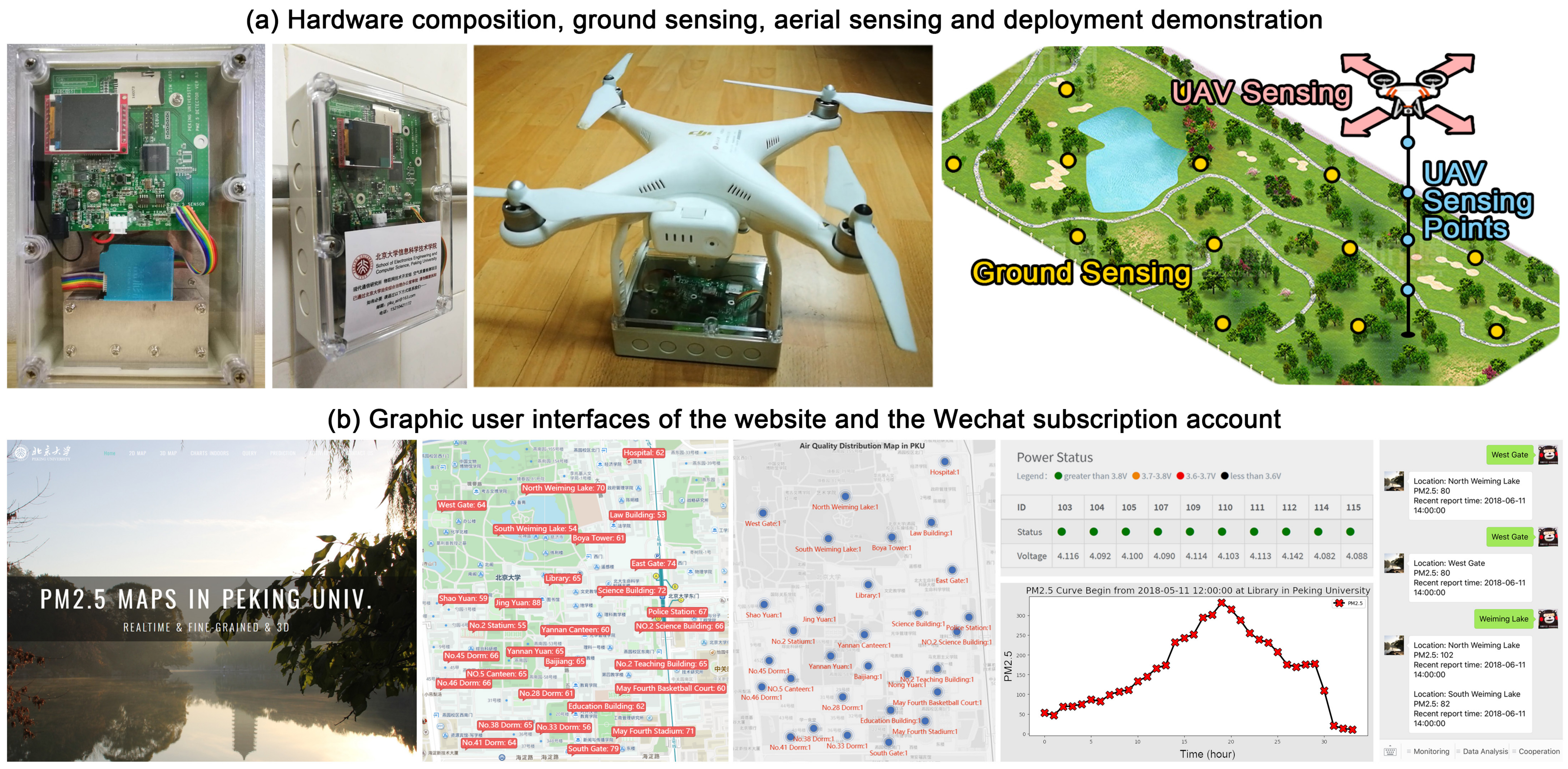}
\vspace{-5mm}
\caption{An illustration of our implementation, where (a) shows the hardware and deployment, (b) shows the user interfaces.}\label{Fig_2_HardwareAndSoftware}
\end{figure}

\emph{}\!\!
The information presented to the users should be well organized.
In our implementation, the users are able to acquire their interested information through their computer or mobile phone on our official website~\cite{bib_Website} and the WeChat official subscription account, shown in Fig.~\ref{Fig_2_HardwareAndSoftware}~(b).
They can see the current distribution of air quality, observe the historical data of certain locations, and check the example of short-term predictions.

\subsubsection{GUI for System Manager}\emph{}\!\!
The information presented to the system manager includes all the raw data and processed data of air quality, as well as the state of the sensing devices such as power percentage and temperature.
The data of air quality can be extracted according to filtering rules for the use of further research.
The state of devices can help the manager decide the plan of equipment maintenance.

\section{Data processing}\label{sec_Processing}

In this section, the most significant functions of the processing layer are presented.
The first subsection introduces the preprocessing of the raw data.
The second subsection discusses the importance of spacial fitting and shows the adopted method.
Finally, the third subsection provides with the solution of short-term prediction of air quality.

\subsection{Preprocessing}


\subsubsection{Preprocessing}\emph{}\!\!
The raw data received by the server in the processing layer may contain unexpected outliers (abnormal values).
There are mainly two kinds of reasons for the appearance of  outliers.
The first one is the unexpected temporary change of the local air quality, which could be caused by the passing-by creatures or vehicles during the detection period.
Such influence is correctly recorded but also affect the result of spacial fitting and short-term prediction.
The second one is the false detection of sensors, which could be caused by the unexpected electrical noise or the possible detecting failure in the long-term and wide-range deployment.
The server recognizes the outliers by analysing each value's relative error compared with the collected values at other locations.
The corresponding values are ignored in the following procedures of spacial fitting and short-term prediction.

\subsection{Spacial Fitting}\label{sec_DataProcessing_SpatialFitting}

Spacial fitting indicates that the server utilizes the current data of measured locations and some of the historical data to provide the best guess for the values at unmeasured locations.
There are two basic proposes of spacial fitting.
First, the missing values on the air quality map can be complemented when some values are eliminated as outliers or a sensing device fails to upload its latest data.
Second, the air quality of the locations where no sensing devices are deployed can be acquired, in which way a even more fine-grained distribution of air quality can be established.

In our implementation~\cite{bib_YuzheINFOCOM}, a machine learning based solution is adopted.
To acquire the data of an unmeasured location at a certain time $D(x,y,z,t)$, a screening process is first executed to choose the most relevant known values which can help determine $D(x,y,z,t)$.
The classic $k$-Nearest Neighbour ($k$NN) algorithm is utilized in the screening process, where both spatial $k$NN and temporal $k$NN are included.
The spatial $k$NN selects $k$ locations with known values, based on spatial Euclidean distance.
And the temporal $k$NN selects $k$ locations with known values, in the condition that these $k$ locations have the most similar temporal change patterns.
Based on the screening results, the historical data of these relevant locations are taken as input of a Deep Neural Network (DNN) to train the possible value of the unmeasured location.
The advantage of DNN is its ability to fit any complicated mapping relations between the input and the output, therefore, the hidden pattern of the spatial distribution of the air quality can be established.
The above screening process takes the advantage of the spatial-temporal correlation of the distribution of air quality, where only the data from relevant locations are considered as valuable inputs.
Therefore, the complexity of the DNN training model can be decreased and a satisfactory precision of the fitting results can be guaranteed.

\begin{figure}[!thp]
\centering
\includegraphics[width=5in]{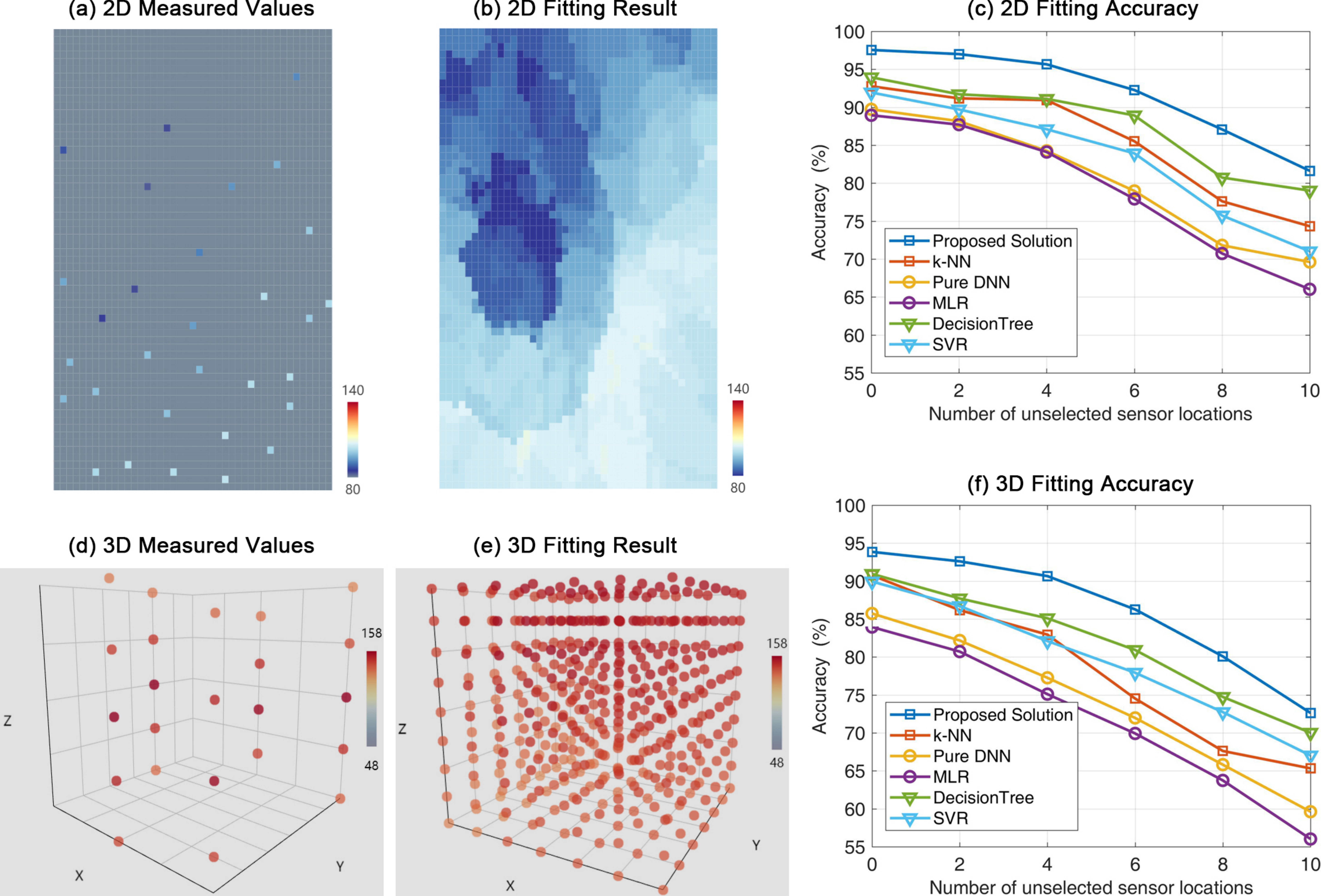}
\vspace{-5mm}
\caption{An illustration of  the spacial fitting in the processing layer, with (a) (b) (c) for 2D fitting and (d) (e) (f) for 3D fitting.}\label{Fig_3_SpatialFitting}
\end{figure}

Fig.~\ref{Fig_3_SpatialFitting} provides an example of spacial fitting.
With limited number of measured points, a fine-grained 2D/3D map can be established.
To test the accuracy, several measured points are deleted from the inputs of the algorithm, used to test the fitting results.
As shown in Fig.~\ref{Fig_3_SpatialFitting} (c) and (f), the proposed solution outperforms the other classical training methods, such as the pure $k$NN, the pure DNN, Multi-variable Linear Regression (MLR) and Support Vector Regression (SVR).

\subsection{Short-Term Prediction}\label{sec_DataProcessing_Prediction}

Short-term prediction means that the server utilizes the historical data and other weather conditions as the training data to determine the possible air quality at different locations in the near future.
There are two basic proposes of prediction.
First, the latency of data uploading that induced by the power control scheme can be accommodated.
The system can provide a predicted air quality distribution based on the historical data, before receiving the real measured values from the sensing devices.
Second, users are able to check the conjectural air quality distribution in the near future, as a reference to determine their behaviours, such as wearing masks or keeping away from the place with higher pollution.

In our implementation~\cite{bib_YuzheINFOCOM}, the adopted solution is also based on the $k$NN and DNN methods just as discussed in Section~\ref{sec_DataProcessing_SpatialFitting}.
However, there are two major improvements compared with the spatial fitting.
The first one is to regard the time dimension as the fourth spatial dimension in the data processing, in which way predicting the future air quality can be seem as fitting the values in the fourth dimension.
The second one is the additional input data as the input of the DNN training procedure.
To provide the prediction values of the future distribution of the air quality at different locations, we have to obtain additional parameters of the weather conditions, such as the wind, the humidity and the temperature.
These parameters can be recorded in our server by scraping data from the online official weather forecast of the meteorological bureau.
In DNN, the hidden influence of the historical data of these weather conditions on the historical data of air quality can be learnt.
Therefore, the air quality in the near future can be predicted.

\begin{figure}[!thp]
\centering
\includegraphics[width=6in]{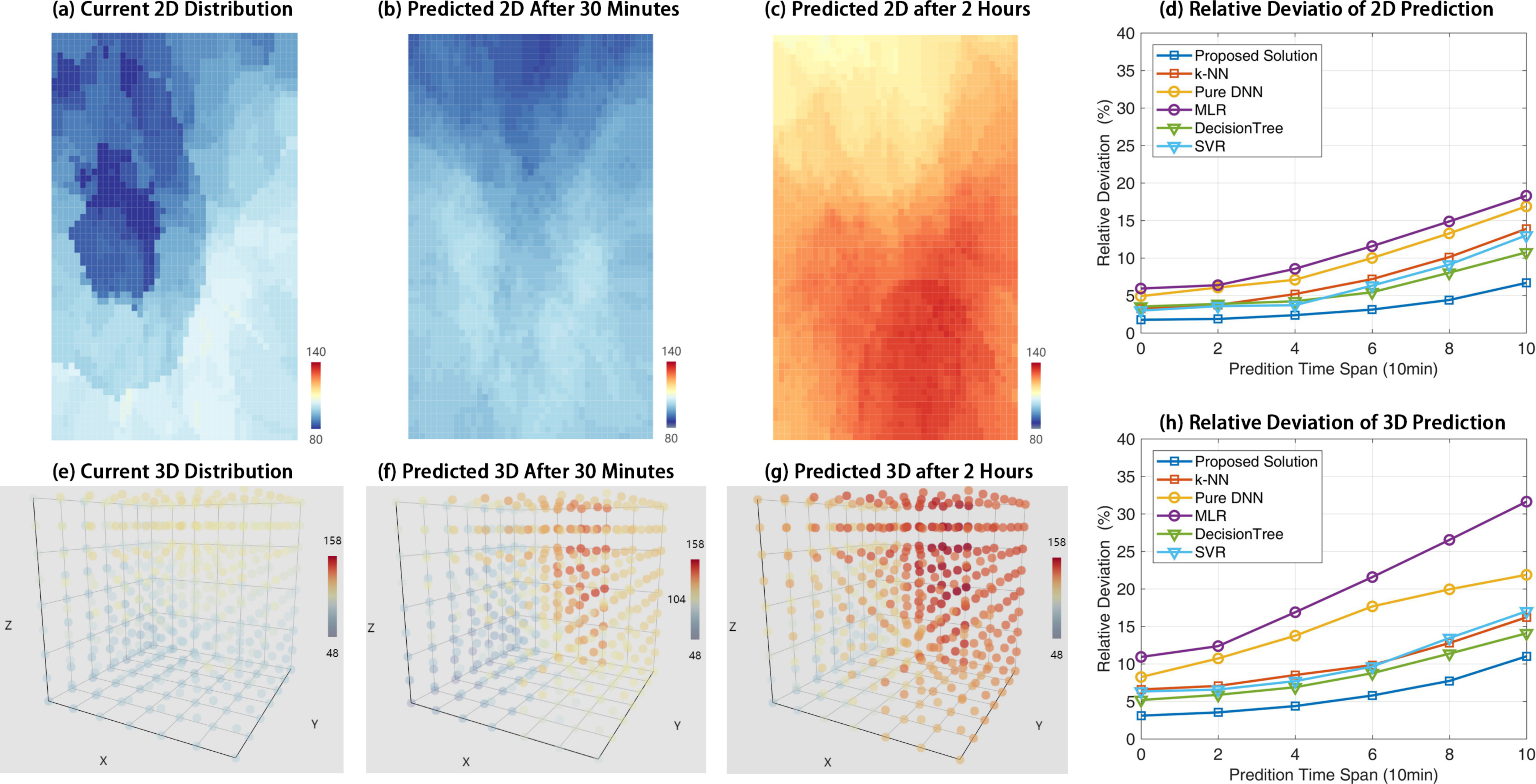}
\vspace{-5mm}
\caption{An illustration of the short-term prediction in the processing layer, with (a) (b) (c) (d) for 2D predicting and (e) (f) (g) (h) for 3D predicting.}\label{Fig_4_ShortTermPrefiction}
\end{figure}

Fig.~\ref{Fig_4_ShortTermPrefiction} provides an example of short-term prediction.
With all the historical data and the spatial fitting results, a short-term prediction can be done for both 2D and 3D scenarios.
The relative deviation of the results of prediction increases with the time span.
The prediction within one hour based on the proposed solution is quite accurate, with only $7\%$ error for 2D prediction and only $11\%$ error for 3D prediction.

\section{Deployment Strategies}\label{sec_Deployment}

In the air quality sensing system, there are limited number of sensing devices on the ground and limited power of the UAV to detect a wide-range 3D space.
The deployment strategies could influence the effectiveness of the spatial fitting and short-term prediction.
For ground sensing, the locations of the fixed sensing devices should be properly selected, so as to establish a reasonable distribution of monitoring points.
For aerial sensing, the aerial sensing positions and routing scheme should be carefully considered, so as to improve the completeness of the spatial detection.
Due to the difference between the problem of ground sensing deployment and the problem of aerial sensing deployment, we discuss them in the following two subsections along with the corresponding solutions.

\subsection{Ground Sensing: Deployment Location Selection}\label{sec_Deployment_Ground}

In order to build a fine-grained air quality map, the technique of spacial fitting is presented in Section~\ref{sec_DataProcessing_SpatialFitting}.
However, the prerequisite of recovering a reliable fine-grained air quality map is that the measured locations should be reasonably distributed.
If most of the sensing devices are deployed only in a small region of the whole concerned area, then the precisions of the sensed data in different regions will not be balanced.
The region with the lowest density of sensing devices could be a weakness of the whole system, since the overall accuracy of the spacial fitting depends on the most inaccurate data among all the input values.
Therefore, a location selection strategy for the ground sensing should be studied.
It is noticeable that such deployment strategy is often based on the collected data from the sensing devices that already been deployed.
In other words, this strategy is actually a strategy to adjust the locations of the ground sensors rather than the initial deployment strategy.

In our implementation~\cite{bib_YuzheICC}, the key idea is to find the deployment strategy that leads to the least uncertainty of the air quality at the unmeasured locations.
We adopt \emph{entropy} to quantitatively describe such uncertainty, which is a physical quantity often used to describe the degree of chaos in physics or the amount of information in informatics.
The entropy of the air quality at a certain unmeasured location depends on the distribution of its possible values.
In general, the wider the distribution is, the higher its entropy will be.
The distribution of the value at an unmeasured location is the linear superposition of the distribution of the values at related locations with different weight.
The weight can be obtained by applying a semi-supervised learning (SSL) method, with the objective to minimize the overall entropy.
When the SSL converges, the entropy of all the unmeasured locations can be obtained.
To select $n$ deployment locations out of $N$ total possible locations, a greedy algorithm can be applied.
First, given an initial set $\{\mathcal{L}\}$ of $n$ selected locations, this algorithm iteratively find one location from the unmeasured locations to replace a selected location in $\{\mathcal{L}\}$.
Each step of replacing should decrease the average entropy of the unmeasured locations, and the iteration stops when the entropy cannot be reduced by replacing one location.

\begin{figure}[!thp]
\centering
\includegraphics[width=5in]{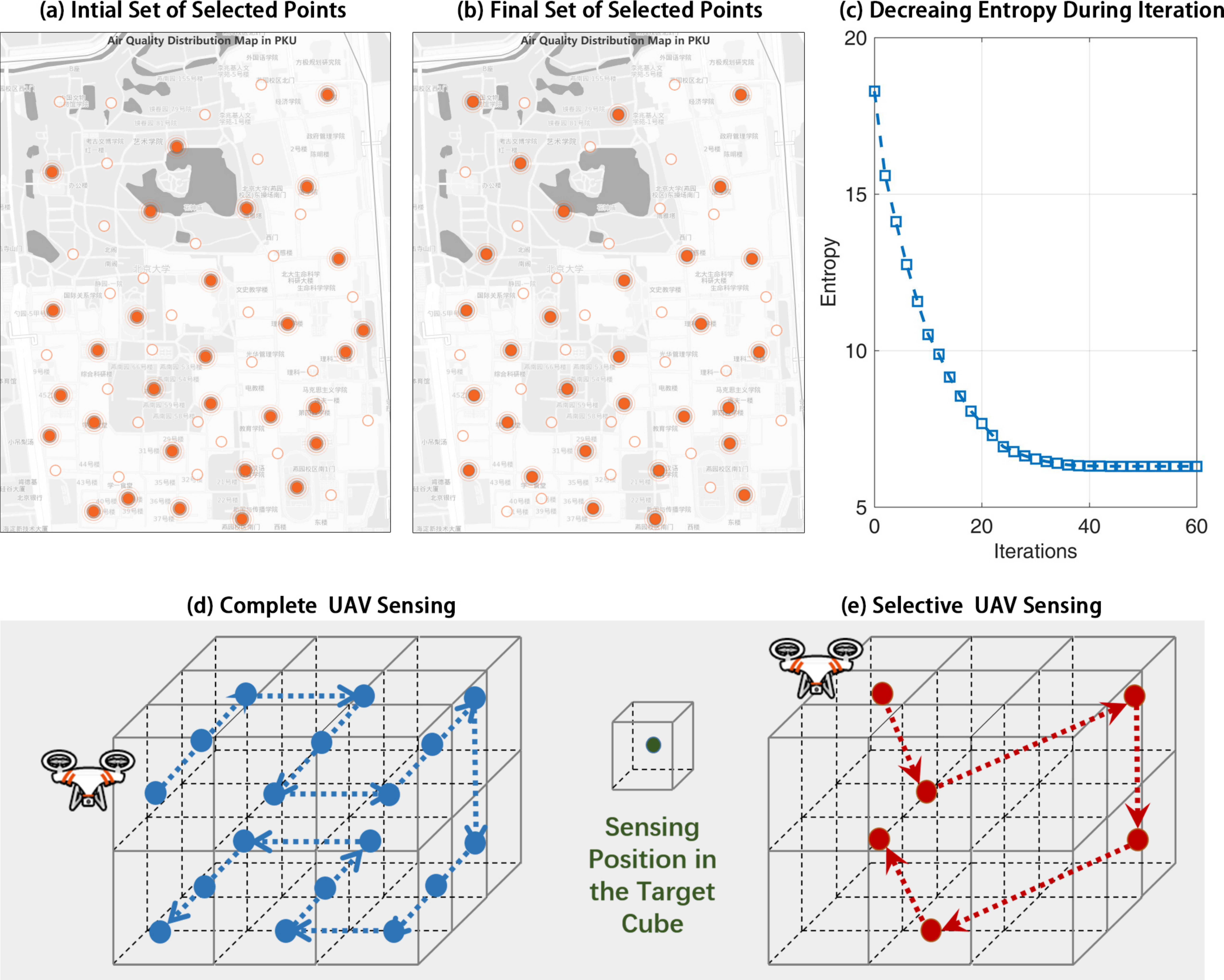}
\vspace{-5mm}
\caption{The deployment strategies, where (a) (b) (c) show the deployment process of ground sensing, (e) (f) explain the difference between complete aerial sensing and selective aerial sensing.}\label{Fig_5_Deployment}
\end{figure}

Fig.~\ref{Fig_5_Deployment} (a) (b) (c) illustrate the procedure of selecting deployment locations of ground sensing.
A better set of locations is found by the minimizing the entropy of unselected points based on the collected data.

\subsection{Aerial sensing: Detecting Location Selection and Routing}

The intrinsic weakness of the small-scale UAVs is their limited power capacity.
As most of the power consumption is caused  by the engine, the operation time of a UAV is highly constrained, usually within only half an hour.
In addition, to detect the air quality at a specific location, the UAV needs to keep stabilized for a few seconds, waiting for the air going through the venting tunnel of the sensing device.
This feature further hinders the attempt to measure the air quality in the whole 3D space.
Therefore, a typical way is to perform a selective sensing routine instead of a complete sensing routine.
The incomplete data of the air quality in the 3D space can be used to recover the complete data by the spacial fitting techniques introduced in Section~\ref{sec_DataProcessing_SpatialFitting}.
A strategy of selecting limited number of aerial positions needs to be investigated, and the flying route of the UAV should also be designed.

In our implementation~\cite{bib_YuzheIOT}, the selection of sensing positions is based on a different scheme compared with the one in Section~\ref{sec_Deployment_Ground}, due to the lack of long-term data that can be collected by the UAV.
The key idea is to first execute a detection for the concerned 3D space as entirely as possible.
Then a simple spacial fitting is performed to create a more detailed distribution map of the air quality.
To select sensing positions, the adopted indicator is called as Partial Derivative Threshold (PDT)~\cite{bib_YuzheGLOBECOM}.
It is defined based on the derivatives of the concentration of the PM value in three spatial dimensions.
Two kinds of positions are considered as important points, namely, the positions with relatively high changing rates, and the positions that contain minimum or maximum values.
Each important point is associated with a corresponding value that indicates its importance.
The routing of the UAV is a NP-hard problem known as the Travelling Salesman Problem (TSP).
Thus, a greedy algorithm is performed to choose the next sensing position at each step.
The criterion is based on the ratio of the importance of a specific point and the cost of movement.
The positions that are selected in this algorithm can be measured by the UAV before its battery runs out.

Fig.~\ref{Fig_5_Deployment} (d) (e) illustrate the selective sensing positions and the routing in aerial sensing.
The space is regarded as a set of cubes, where the only a proportion of them are added in the aerial sensing route.

\section{Power Control}\label{sec_Power}

For the sensing system, there is a intrinsic tradeoff between the power consumption and the precision.
To be specific, the precision here refers to the precision of the established fine-grained spatial-temporal air quality distribution based on the measured data.
A more frequent data collection procedure can provide a better continuity and a better precision, but it also induces additional power consumption.
In the following two subsections, the sensing/uploading intervals of the ground sensing and the flight plan of the aerial sensing are discussed.

\subsection{Ground Sensing: Sensing Intervals and Uploading Intervals}

For the sensing devices without external power supplies, the operations of sensing and uploading contribute nearly $85\%$ of the power consumption.
If a ground sensing device has a short sensing interval and a short uploading interval, then the duration of its battery will not be very long.
As soon as most of the deployed sensing devices run out of electricity, the function of the system has to be paused and the devices need to be retrieved, recharged and re-deployed.
If a ground sensing device has a long sensing interval or a long uploading interval, the the server has to do more spatial fitting or short-term predictions in order to keep the displayed data to be real-time.
The longer the sensing/uploading intervals are, the lower the accuracy of the fitting and prediction results will be.

\begin{figure}[!thp]
\centering
\includegraphics[width=6in]{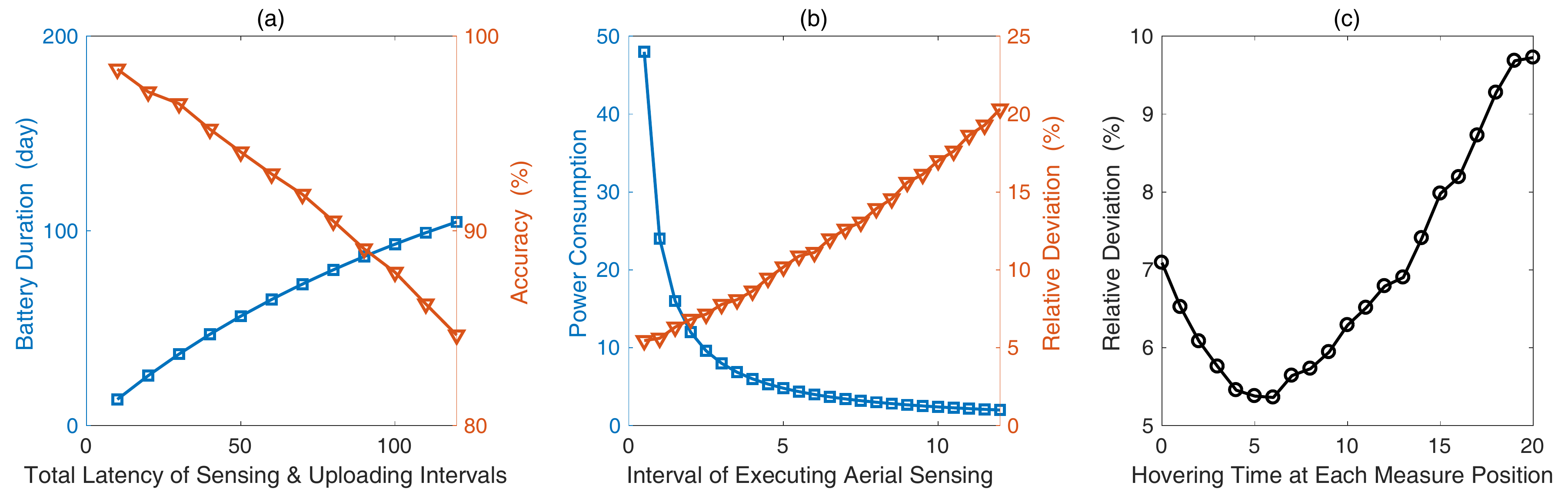}
\vspace{-5mm}
\caption{The power control tradeoffs, where (a) is for ground sensing, (b) and (c) are for aerial sensing.}\label{Fig_6_PowerControl}
\end{figure}

Fig.~\ref{Fig_6_PowerControl}~(a) shows the tradeoff.
As the latency induced by sensing and uploading intervals increases, the battery duration increases but the average accuracy decreases.
In our implementation, the setting is based on practical considerations.
The default sensing interval is 30min and uploading interval is 60min.
Shorter intervals are being applied when the air quality is getting worse.

\subsection{Aerial sensing: Flight Plan}

For aerial sensing, the major constraint is from the power-limited UAV rather than the sensing device that tied beneath.
After each time of flight, the low-electricity battery of the UAV has to be replaced by a full-electricity battery, so as to continue the next round of aerial sensing.
The cost of each flight is high, and it is not expected that aerial sensing can be executed all the time.
During the period without UAV sensing, the server has to utilize the pattern that it learnt from the spatial-temporal historical data that it has collects.
Therefore, the number of flight that is done in one day should be determined to balance between the cost and the precision.
Another controllable factor of the UAV flight plan is the hovering time that the UAV stays at each selected position.
Due to the limited ventilation capacity of the sensing device, the UAV has to stay for a while at each position to make sure that the collected data is indeed based on the air at this position instead of the residual air from the previous location.
A long hovering time can guarantee the accuracy of data collecting but also consume additional power of the UAV, which in turn reduces the number of positions that the UAV can sense.

Fig.~\ref{Fig_6_PowerControl}~(b) shows the influence of the aerial sensing interval.
As the sensing interval gets larger, the total power consumption decreases significantly but the relative deviation of fitting and prediction gets larger.
In our implementation, the default aerial sensing interval is $6$ to $12$ hours.
Fig.~\ref{Fig_6_PowerControl}~(c) shows the influence of the UAV hovering time at each measure position.
The best decision of UAV hovering time is around 5 seconds, to balance between the error of air sampling and the number of sensing positions.

\section{Conclusion}\label{sec_Conclusion}
In this article, an IoT-based 3D air quality sensing system has been presented, which is designed to be fine-grained, real-time, and power efficient.
The architecture of this system includes the sensing layer to collect data, the transmission layer to enable bidirectional communications, the processing layer to analyze and process the data, and the presentation layer to provide graphic interface for users.
In the sensing layer, the sensing devices are not expected to sense and upload data continuously, due to their limited battery capacity.
In the processing layer, spatial fitting is done by the server so as to establish the fine-grained air quality distribution with limited number of sensing devices.
Additionally, short-term prediction provides real-time air quality distribution in the face of data latency.
To further optimize the quality of the collected data, deployment strategies for ground sensing and aerial sensing are executed to adjust the sensing locations.
At last, the power control for ground sensing and aerial sensing are considered, to balance between the data accuracy and the power consumption.



\begin{thebibliography}{15}

\bibitem{bib_WHO}
World Health Organization, ``7 Million Premature Deaths Annually Linked to Air Pollution," \emph{Air Quality Climate Change}, vol.~22, no.~1, pp.~53-59, Mar.~2014.

\bibitem{bib_AriPollution}
Q.~Di, Y.~Wang, A.~Zanobetti, et al, ``Air Pollution and Mortality in the Medicare Population,'' \emph{New England J. of Medicine}, vol.~376, no.~26, pp.~2513-2522. Jul.~2017.

\bibitem{bib_AQI}
G.~Kyrkilis, A.~Chaloulakou, and P.~A.~Kassomenos, ``Development of an Aggregate Air Quality Index for an Urban Mediterranean Agglomeration: Relation to Potential Health Effects", \emph{Environment International}, vol.~33, no.~5, pp.~670-676, 2007.

\bibitem{bib_Sensing}
Y.~Gao, W.~Dong, K.~Guo, X.~Liu, Y.~Chen, X.~Liu, J.~Bu and C.~Chen, ``Mosaic: A Low-Cost Mobile Sensing System for Urban Air Quality Monitoring,'' \emph{IEEE International Conference on Computer Communications}, San Francisco, CA, Jul. 2016.

\bibitem{bib_Station}
Beijing MEMC. (Mar. 2017). \emph{Beijing Municipal Environmental Monitoring Center}. [Online]. Available: http://www.bjmemc.com.cn/

\bibitem{bib_Meter2Meter}
T.~Quang et al., ``Vertical Particle Concentration Profiles Around Urban Office Buildings," \emph{Atmospheric Chemistry and Physics}, vol.~12, no.~11, pp.~5017-5030, May~2012.

\bibitem{bib_IoTAir}
W.~Fuertes, D.~Carrera, C.~Villac\'{\i}s, T.~Toulkeridis, F.~Gal\'{a}rraga, E.~Torres, and H.~Aules, ``Distributed System as Internet of Things for a New Low-Cost, Air Pollution Wireless Monitoring on Real Time," in \emph{Proc. IEEE/ACM 19th International Symposium on Distributed Simulation and Real Time Applications}, Chengdu, China, Oct.~2015.

\bibitem{bib_IoTData}
R.~Ranjan, O.~Rana, S.~Nepal, et al, ``The Next Grand Challenges: Integrating the Internet of Things and Data Science," \emph{IEEE Cloud Computing}, vol.~5, no.~3, pp.~12-26, june~2018.

\bibitem{bib_Suggestions}
C.~Borrego, H.~Martins, O.~Tchepel,	 L.~Salmim, A.~Monteiro, and A.~I.~Miranda, ``How Urban Structure can Affect City Sustainability from an Air Quality Perspective," \emph{Environmental modelling \& software}, vol.~21, no.~4, pp.~461-467, Apr.~2006.

\bibitem{bib_DataSet}
Dataset Collected from Peking University and Xidian University. [Online]. Available: https://github.com/pku-bzx/pku-air.

\bibitem{bib_AdHoc}
K.~Tatara, G.~Lee, and N.~Y.~Chong, ``Self-Organizing Ad-hoc Robotic Sensor Networks Based on Locally Communicative Interactions," in \emph{Proc. International Conference on Ubiquitous Robots and Ambient Intelligence}, Incheon, South Korea, Nov.~2011.

\bibitem{bib_Website}
Our Website-based GUI. [Online]. Available: http://www.aqimaps.com.

\bibitem{bib_YuzheINFOCOM}
Y.~Yang, Z.~Bai, Z.~Hu, Z.~Zheng, K.~Bian, and L.~Song, ``AQNet: Fine-Grained 3D Spatio-Temporal Air Quality Monitoring by Aerial-Ground WSN", in \emph{Proc. IEEE  International Conference on Computer Communications}, Honolulu, USA, Apr.~2018.

\bibitem{bib_YuzheIOT}
Y.~Yang, Z.~Zheng,  Y.~Jiang, L.~Song, and Z.~Han, ``Real-Time Profiling of Fine-Grained Air Quality Index Distribution Using UAV Sensing," \emph{IEEE Internet of Things}, vol.~5, no.~1, pp.~186-198, Feb.~2018.

\bibitem{bib_YuzheGLOBECOM}
Y.~Yang, Z.~Zheng, K.~Bian, Y.~Jiang, L.~Song, and Z.~Han, ``Arms: A Fine-Grained 3D AQI Realtime Monitoring System by UAV," in \emph{Proc. IEEE Global Communications Conference}, Singapore, Dec.~2017.

\bibitem{bib_YuzheICC}
Y.~Yang, Z.~Zheng, K.~Bian, L.~Song, and Z.~Han, ``Sensor Deployment Recommendation for 3D Fine-Grained Air Quality Monitoring using Semi-Supervised Learning," in \emph{Proc. IEEE International Conference on Communications}, Kansas, USA, May 2018.


\end{thebibliography}
\end{document}